\documentclass[mathleft
]{an}
\usepackage{graphicx}
\usepackage{times}
\usepackage{color}
\newcommand{\sr}{R_{\odot}}

\begin{document}

% The following seven commands are intended for editorial usage and
% should be ignored by
% the author(s).
\Pagespan{1}{}% Document's page range.
% If second parameter is left empty, the last page is computed
% automatically.
\Yearpublication{}%
\Yearsubmission{}%
\Month{}%
\Volume{}%
\Issue{}%
% \DOI{This.is/not.aDOI}%

\title{Flux-dominated solar dynamo model with a thin shear layer}

\author{G.A. Guerrero\inst{1}\fnmsep\thanks{Corresponding author:
  \email{guerrero@astro.iag.usp.br}\newline}
%Example
%for footnote, note the usage of the \texttt{fnmsep}
%command as separator between institute number and footnote mark}
\and  E.M. de Gouveia Dal Pino\inst{1} }
\titlerunning{Tachocline thickness}
\authorrunning{G. Guerrero \& E.M. de Gouveia Dal Pino}
\institute{
Instituto Astronomico e Geof\'{i}sico, IAG, USP,
Rua do Mat\~{a}o, 1226, S\~{a}o Paulo, Brasil
}

\received{01 Aug 2007}
\accepted{}
\publonline{later}

\abstract{ Flux-dominated solar dynamo models have demonstrated to
reproduce the main features of the large scale solar magnetic cycle,
however the use of a solar like differential rotation profile
implies in the the formation of strong toroidal magnetic fields at
high latitudes where they are not observed. In this work, we invoke
the hypothesis of a thin-width tachocline in order to confine the
high-latitude toroidal magnetic fields to a small area below the
overshoot layer, thus avoiding its influence on a Babcock-Leighton
type dynamo process. Our results favor a dynamo operating inside the
convection zone with a tachocline that essentially works as a
storage region when it coincides with the overshoot layer. }

\maketitle
\section{Introduction}
Flux-dominated solar dynamo models, which use a solar like velocity
field, together with an estimated diffusivity \linebreak profile and
an $\alpha$ effect resembling the Babcock-Leighton mechanism for
producing poloidal magnetic fields, have \linebreak  demonstrated to provide
solutions that resemble quite well the observations of the large scale
solar magnetic cycle \linebreak (Dikpati et al. 2004, Chatterjee, Nandy \&
Choudhuri 2004, Guerrero \& de Gouveia Dal Pino 2007, hereafter GDP).

In this kind of models, the usually accepted scenario for the dynamo
operation can be described as follows: starting with a dipolar
field, the $\Omega$ effect takes place both at the convection zone
and the tachocline, and a toroidal magnetic field is formed and
amplified by the radial and latitudinal components of the
differential rotation. Once this field reaches values between $10^4$
- $10^5$ G, it becomes buoyant unstable and arises through the
convection zone being twisted by the Coriolis force until it emerges
at the surface and forms bipolar magnetic regions. This part of the
process is known as Babcock - Leighton $\alpha$ effect. At the
surface, the leading part of a bipolar region points to the equator,
while the rear part is drifted polarward by the meridional
circulation. This migration contributes to the formation of big
loops in each hemisphere, which later will reconnect in order to
form a new poloidal field of opposite polarity to the initial one.

In view of the present status of the
observations, the model proposed above presents several problems, as
\linebreak remarked by Brandenburg (2005), the most important of which
is related to the location, the sign and the amplitude of the alpha
effect. In the scenario discussed above, the $\alpha$ term
should  reproduce the emergence of magnetic flux tubes of strong
toroidal magnetic fields and then the migration of these fields
toward the poles. In this way, the most probable location for the
dominance of this
term would be the upper layers of the convection zone, with a
latitudinal  distribution peaking around $30 ^{\circ}$, which is the
latitude at which the strongest magnetic activity is observed. If
the solar dynamo is operating at the tachocline, where the shear has
positive values, the natural sign of the alpha effect should be
negative (positive) in the north (south) hemisphere, in order to get
equatorward branches in agreement with the mean field dynamo theory.
However, the presence of an equartorward meridional flow at deeper
layers could bring the possibility of a positive (negative) $\alpha$
effect at the north (south) hemisphere and still give the correct
migration. Recent works have suggested that the amplitude of the
$\alpha$ term could be important in the determination of the parity of
the dynamo (Dikpati \& Gilman 2001, Bonnano et al. 2002), this effect is
not considered in this work since our domain corresponds to one
hemisphere only, but it will be addressed in a forthcoming paper
(Guerrero \& de Gouveia Dal Pino 2007, in preparation).

Another important question regards the real importance of the
tachocline in the dynamo process. As a layer of strong radial shear,
the tachocline could be the ideal site for the amplification of the
magnetic field, though the maximum amplitude of  the shear is
located at high latitudes, which suggests an intense activity  close
to the poles. In this case, meridional circulation could be again an
appropriate solution to this problem, since a velocity of about $3$
m s$^{-1}$ could advect down and equatorward the magnetic field
across a stable layer until reaching the latitudes where the strong
activity is observed (Nandy \& Choudhuri 2002, Chatterjee et al.
2004). However, the use of a deep meridional  flow in a flux
dominated dynamo has been found to be very sensitive to the adopted
$\alpha$ effect and  the diffusion and velocity profiles (Guerrero
\& Mu\~{n}oz 2004) and can also generate undesirable abundance
variations in the convection zone. In addition, several numerical
simulations (Gilman \& Miesh 2004, R\"{u}diger, Kitchatinov \& Arlt
2005) have demonstrated the impossibility of penetration of the
meridional flow below the tachocline more than a few kilometers. In
order to allow some shallower penetration, other possibilities, such
as overshooting or turbulent magnetic pumping have been invoked in the 
literature (Rogers, Glatzmaier \& Jones 2006, Ziegler \& R\"{u}diger
2003, respectively), where magnetic flux tubes are drifted
to the overshoot layer and can be amplified to the desired
magnitudes.

Indeed, if a flow penetrates inside the tachocline, there should be
a mechanism that could prevent either the formation of strong
toroidal magnetic fields at higher latitudes or the participation of
these fields in the cycle which is responsible for the observed
phenomena. In recent work (GDP), we have shown that a possible
solution to the problem above could be the choice of an appropriate
set of physical parameters, since the model is sensitive to the
location of the transition between the sub-adiabatic and the
super-adiabatic layers, to the diffusivity value at the bulk of the
convection zone and to the thickness of the solar tachocline (see
Figs. 8a and 8b of GDP). We have also found that the latitudinal
shear term is dominant over the radial shear term throughout the
cycle, and that the radial shear peaks only in the early years of
the cycle and decays afterward  to values about two orders of
magnitude smaller than the latitudinal component.

In this paper, we present a detailed description on how the
suppression of the magnetic fields at high latitudes is obtained. We
also present the results when two different prescriptions for the
meridional circulation profile are employed. They seem to be
insensitive to these variations.

\section{Model}

Our model is quite similar to the one explained in detail in
Guerrero \& Mu\~{n}oz 2004 and GDP. In essence, we solve the mean
field induction equation in the $r$ and $\theta$
coordinates by using a second order finite difference scheme for the
spatial derivatives (Lax-Wendroff for the first derivatives and FTCS
for the second ones, Press et al. 2002) and the second order ADI
method for the temporal advance \footnote{For sake of
    simplicity we have not included in our induction equation a
    diamagnetic diffusion term which could be important when gradients 
    of turbulent diffusivity are present (see e.g., Kitchatinov \&
    R\"{u}diger, 1992)}. For the present 
work, we have written a new parallel version of the code using the MPI
libraries, obtaining a speedup factor around $3$ in a four-processor
computer. This has allowed us to use a resolution of 200 x 200 grid
points, with an adjustable time step to fulfill the Courant condition
and perform long and fast numerical simulations.

\begin{table}
\centering%%%
\caption{Parameters employed in the present model compared to those of
  GDP. $U_0$ is the maximum amplitude of the meridional circulation at
  the surface, $R_p$ is the penetration depth of the flow, $\beta_1$
  and $\beta_2$ fit the latitudinal profile of the meridional
  circulation to the helioseismology results, $\eta_c$ and $\eta_s$
  are the values of the magnetic diffusivity at the convection zone
  and at the surface, respectively.}
\label{tlab}
\begin{tabular}{ccc}\hline
Parameter & Value in GDP & Present value \\ \hline
$U_0$ & $25$ m s$^{-1}$ & $10$ m s$^{-1}$ \\
$R_p$ & $0.69 \sr$ & $0.70 \sr$ \\
$\beta_1$ & $6.06 \times 10^9$ cm$^{-1}$ & $3.47 \times 10^{10}$
cm$^{-1}$\\
$\beta_2$ & $4.6 \times 10^9$ cm$^{-1}$ & $1.39 \times 10^{10}$
cm$^{-1}$\\
\hline
$\eta_c$  & $5\times 10^9$cm$^2$s$^{-1}$ & $3\times 10^{10}
$cm$^2$s$^{-1}$ \\
$\eta_s$  & $1\times 10^{12}$cm$^2$s$^{-1}$ & $3\times 10^{12}
$cm$^2$s$^{-1}$ \\
\hline
\end{tabular}
\end{table}

\section{Results}

\begin{figure*}
\includegraphics[scale=0.72]{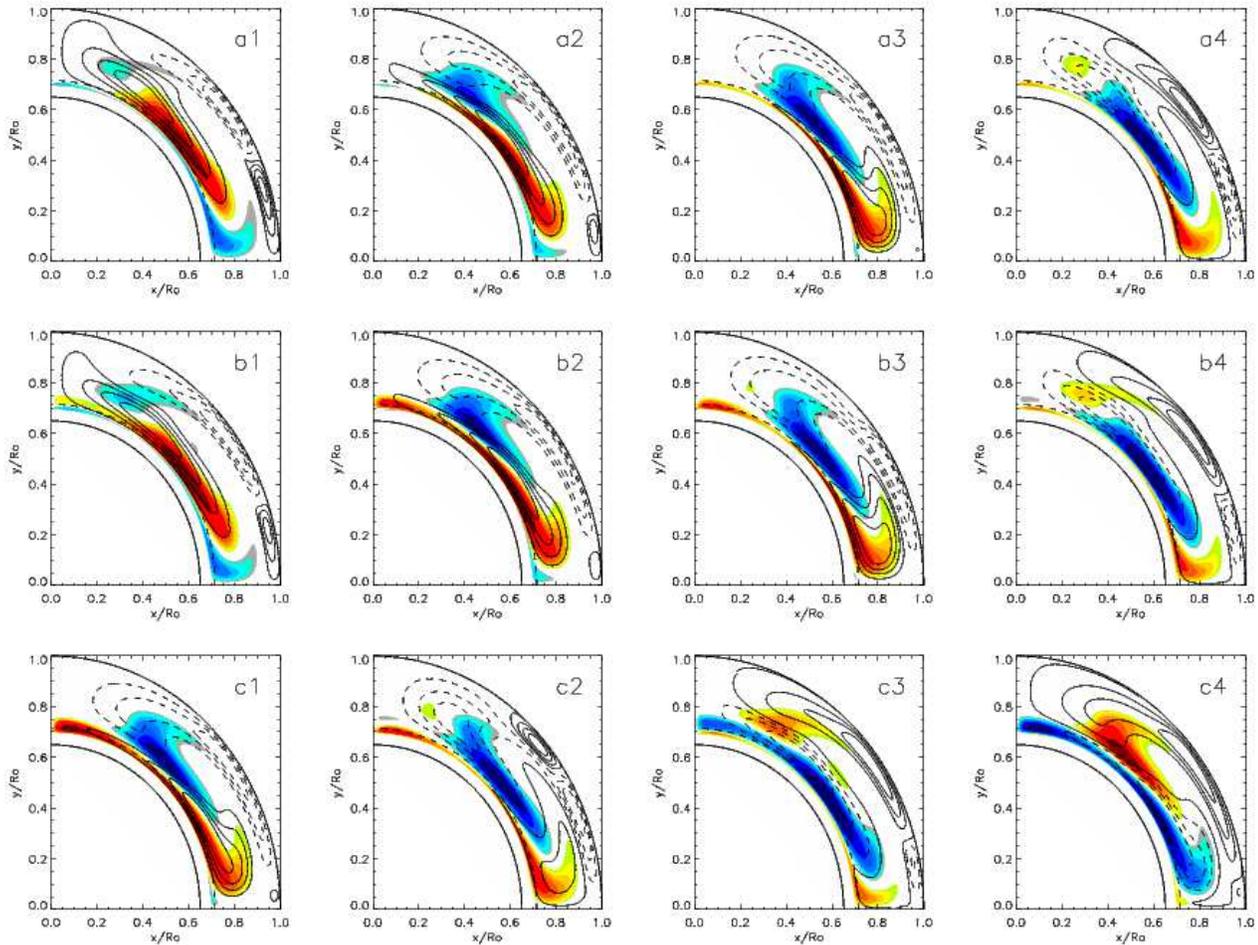}\\
\caption{Snapshots of the positive (negative) toroidal field contours
  are shown in blue (red) scale together with the field lines of the
  positive (negative) poloidal field in continuous (dashed) lines,
  for four different times ($T/8$, $T/4$, $3 T /8$ and $T/2$) along a
  half 11-year cycle. The upper (a), middle
  (b) and bottom (c) panels present the same snapshots for models
  with a thin ($0.02 \sr$), intermediate ($0.06 \sr$) and thick ($0.1
  \sr$) tachocline.}
\label{label1}
\end{figure*}

For comparison, we have considered the same velocity and diffusion
profiles of GDP, but have altered the value of some of the
parameters, as indicated in Table 1. We note that, as
in GDP (see Fig. 3 of that work) we have considered two different
profiles for the magnetic diffusivity in the convective zone. In a
thin layer near the top, we have assumed  a supergranular diffusivity,
while in the inner region of the  convective zone  we have adopted a
lower value. Indeed, there is no physical reason why this value should
be smaller than that near the top. However, if we take the same large
value ($10^{12}$ cm$^2$ s$^{-1}$) for the entire convection zone, the
dynamo will enter in the diffusion dominated regime.
Another important alteration is the location of the
eruption of the magnetic flux tubes. In GDP, the toroidal magnetic
fields are transported from the top of the tachocline, i.e. $\alpha$
$\propto$ $B_{\phi}(R_c + \frac{d_1}{2},\theta)$ (where $B_{\phi}$ is
the azimuthal component of the magnetic field, and $R_c$ is the
location of the center of the \linebreak tachocline of thickness
$d_1$), while in the present 
work the eruption point is placed in the middle of the overshoot layer
($r_c$ $=$ $0.715 \sr$, see the dashed line in the snapshots of
Fig. 1). Besides being more realistic, this assumption allows us to
make direct comparisons with the stability analysis of magnetic flux 
tubes in the overshoot layer (see Fig. 1 of Ferriz-Mas \&
Sch\"{u}ssler 1994). \footnote{We note that this scheme is slightly
different from the one
  assumed by Dikpati et. al. (2004), who take the average value for the
  toroidal magnetic field along the overshoot layer.}
Fig. 1 displays four snapshots of the evolution of the toroidal
magnetic field (in color-scale)  and poloidal field (lines) for four
different stages of a half (11-year) cycle. The top, middle and
bottom panels correspond to a thin, intermediate and a thick
tachocline, respectively. As it can be seen, at the beginning of each
new cycle, toroidal magnetic fields start to be formed at two
different places: one part inside of the convective layer,
around $70$ degrees, where the poloidal shear term has its maximum
value; and another one at the tachocline, very close to the poles
where the radial shear reaches its maximum amplitude.

When a thin tachocline is considered (top of Fig. 1), the toroidal
field generated at the poles remains confined below the center of
the overshoot layer.  At the place where we consider the eruption of
the magnetic flux tubes, this high latitude toroidal field does not
reach the necessary values to become buoyant. This effect can be
seen also in the upper panel of  Fig. 2, where a time-latitude
butterfly diagram for the contours of both, the toroidal magnetic
field at $r_c$ (lines) and the radial magnetic field (grey scale
background) at the surface are shown. The contours for the toroidal
field are equally log-spaced for amplitudes above $5 \times 10^4$ G,
which is the value at which the buoyancy begins to develop. Along
with the confinement, we can note that the high latitude toroidal
field diffuses very efficiently since its generation stops, as
quickly as, the radial shear component decreases. On the other hand,
the field which is developed inside the convection zone is advected
down and equartorward by the meridional flow, subtly penetrating the
tachocline and reaching values above $5  \times 10^4$ G .  The
snapshots show that when the toroidal fields are going downward, the
toroidal magnetic field for the new cycle begins to be formed. The
appearance of this new field seems to dislocate the older one and
cause an additional push downward.

\begin{figure}[h]
\includegraphics[scale=0.78]{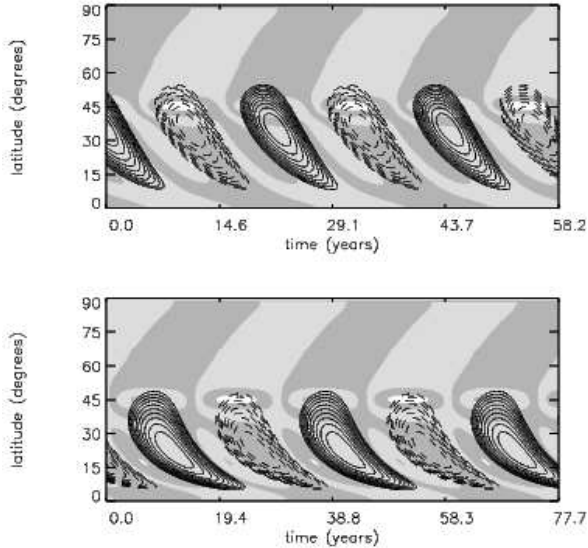}
\caption{Time-latitude butterfly diagram for the model of Table 1 and
  top of Fig. 1 with a thin tachocline (top panel), and for the
  meridional circulation profile used by Dikpati \&
  Charbonneau 1999 (bottom panel). The continuous (dashed) lines
  represent the positive (negative) strength of the toroidal field at
  the center of the overshoot layer ($r_c$$=$$0.715 \sr$). The lines
  are log-spaced and cover the interval between $5 \times 10^4$ -
  $10^5$  G. The background gray scale represents the positive (dark)
  and negative (clear) radial field at the surface. }
\label{label1}
\end{figure}

The middle (b) and bottom (c) panels of Fig. 1 show the same stages
of the cycle for models with an intermediate and a thicker
tachocline. These diagrams show that the radial shear, in spite of
having a smaller amplitude than in the previous case, is distributed
along a larger area, so that a larger portion of magnetic field is
amplified leading to a strong solar activity at high latitudes which
is incompatible  with the observations. Note that this result
differs from the one previously obtained in GDP due to the distinct
location of the domain of influence of the $\alpha$ effect. When the
buoyancy point is placed in the center of the overshoot layer, as in
the present case, only a thin tachocline is able to lead to solar
like results (top panel of Fig. 2).

We have  also tested  the validity of this mechanism to avoid the
formation of high latitude toroidal fields with another meridional
circulation profile. The bottom panel of Fig. 2 shows the butterfly
diagram obtained with the analytical profile used by Dikpati \&
Charbonneau (1999). We note that it reproduces quite well the
observations, with the butterfly wings of the toroidal field only
below $\sim 45$ degrees and with a half cycle period of $15$ years.

Finally, we have also performed simulations similar to those of Fig.
2, but (artificially) turning off the radial shear term (i.e.
$\partial \Omega / \partial r$$=$$0$) in the induction equation
(see, e.g. equations 1 and 2 of GDP) and found that the latitudinal
distribution of the toroidal magnetic fields remains unaltered in
the butterfly diagrams. This result indicates that the fields
generated in the tachocline do not significantly contribute to the
observed activity, suggesting  that in a flux dominated solar dynamo
process with a Babcock-Leighton $\alpha$ effect the real role of the
tachocline is to be, essentially, a storage region, provided that it
coincides with the overshoot layer. This is in agreement with the
conclusions of Dikpati, Gilman \& MacGregor (2005).

\section{Conclusions}

We have described a physical scenario in which a simple
flux-dominated solar dynamo model with an appropriate set of
parameters produces results which are in good agreement with the
observations. The main assumptions and results are summarized below:

\begin{enumerate}
\item
Since an analysis of the evolution of the maximum of the shear terms
along the cycle shows that the latitudinal shear is always dominant
over the radial one (which attains appreciable  values only during
the first years and quickly decays afterward), a possible mechanism
to prevent the formation of toroidal magnetic fields  by the radial
shear at high latitudes could be the  consideration of a thin
tachocline. In this case, we find that  the dynamo will form
a thin layer of toroidal field below the overshoot layer  with
insufficient amplitude to produce buoyant flux tubes. Several
numerical tests have demonstrated the robustness of this assumption
in order to avoid the appearance of sunspots near the poles.
\item
The bulk of the toroidal magnetic field is formed at the interior of
the convection zone thanks to the latitudinal shear and is drifted
by convection and/or magnetic pumping towards the sub-adiabatic
overshoot layer \linebreak where we assume that it
undergoes magnetic buoyancy and rises to the surface at values above
of $5$$\times$$10^4$ G.
Note that the turbulent diamagnetism (which
was mentioned in the section 2) may also play an important role in
order to pile-up the toroidal magnetic field in the overshoot layer,
since it behaves like a downward velocity term in the induction
equation, as demonstrated by \linebreak Kitchatinov \& R\"{u}diger
(1992). The importance of this term has not yet been studied in
detail, but will be considered in forthcoming studies.
\item
The above mentioned results suggest  that the dynamo which is
responsible for the observed activity does not operate at the
tachocline, as usually assumed, but inside the convection zone and
is due to the strong latitudinal shear. In this sense, the real role
of the tachocline, if it coincides with the overshoot
layer, is to store the magnetic flux tubes until they reach the
necessary amplification to become buoyantly unstable.
\end{enumerate}

\acknowledgements We would like to thank to Dr. G. R\"{u}diger for
the invitation to participate of the very interesting Thinkshop in
Postdam and to an anonymous referee for helpful comments on the
manuscript. We also acknowledge partial support from the Brazilian
Agencies CNPq and FAPESP. 

%\newpage%%%%%%%%%%%%%%%%%%%%%%%%%%%%%%%%%%%%%%%%%%%%%%%%%%%%%%

\end{document}